\documentclass[12pt]{article}
\usepackage{sbc-template}

\usepackage[utf8]{inputenc}
\usepackage[brazilian]{babel}
\usepackage{amsmath,amsfonts,amssymb}
\usepackage{xcolor}
\usepackage{url}
\newcommand{\urls}[1]{{\small \url{#1}}}

\usepackage{hyperref} 
\usepackage{graphicx}
\usepackage{enumitem}  
\usepackage[]{quoting}

\makeatletter
\newcommand{\osetNormal}[2]{%
  {\mathop{#2}\limits^{\vbox to 3.5\ex@{\kern-\tw@\ex@
   \hbox{\scriptsize #1}\vss}}}}
\makeatother

\newcommand{\draw}{\mathtt{Draw}}
\newcommand{\drawid}{\mathtt{DID}}
\newcommand{\drawnValue}{d}

\newcommand{\drawSet}{\Delta}

\newcommand{\procid}{\mathtt{PID}}
\newcommand{\procinfo}{\mathtt{info}}
\newcommand{\cnt}{\mathtt{cnt}}

\newcommand{\procStakeholders}{S}
\newcommand{\stakeholder}{s}

\newcommand{\procEligible}{E}
\newcommand{\eligible}{e}

\newcommand{\pk}{{pk}}
\newcommand{\sk}{{sk}}

\newcommand{\prob}[1]{P(#1)}

\newcommand{\share}{{share}}
\newcommand{\mask}{{mask}}

\newcommand{\commit}{C}
\newcommand{\sig}{\sigma}

\newcommand{\hash}{\mathcal{H}}
\newcommand{\sign}[2]{\mathcal{S}(#1, #2)} 
\newcommand{\verif}[3]{\mathcal{V}(#1, #2, #3)} 

\newcommand{\seclevel}{\lambda}
\newcommand{\len}[1]{{|{#1}|}}
\newcommand{\bitset}[1]{\{0,1\}^{#1}}
\newcommand{\samples}{\stackrel{\,_\$}{\gets}}
\newcommand{\cat}{{||}}
\newcommand{\iseq}{\osetNormal{?}{=}}
\DeclareMathOperator{\lcm}{lcm}

\title{A Fair, Traceable, Auditable and Participatory  Randomization Tool for Legal Systems}
\author{Marcos Vinicius M. Silva\inst{1}, \ 
Marcos Antonio Simplicio Jr.\inst{1}, \\ 
Roberto Augusto Castellanos Pfeiffer\inst{2}, \ 
Julio Michael Stern\inst{3}}

\address{ 
 Escola Politecnica, Universidade de S{\~a}o Paulo 
\nextinstitute 
 Law School, Universidade de S{\~a}o Paulo 
\nextinstitute
  Institute of Mathematics and Statistics, Universidade de S{\~a}o Paulo   
 \email{mvsilva@larc.usp.br, mjunior@larc.usp.br,} 
 \vspace{-2mm} 
 \email{roberto.pfeiffer@usp.br,  jstern@ime.usp.br}
}

\date{April 2019}

\begin{document}

\maketitle

\begin{abstract}
Many real-world scenarios require the random selection of one or more individuals from a pool of eligible candidates.
One example of especial social relevance refers to the legal system, in which the jurors and judges are commonly picked according to some probability distribution aiming to avoid biased decisions.
In this scenario, ensuring auditability of the random drawing procedure is imperative to promote confidence in its fairness.
With this goal in mind, this article describes a protocol for random drawings specially designed for use in legal systems.
The proposed design combines the following properties: security by design, ensuring the fairness of the random draw as long as at least one participant behaves honestly; 
auditability by any interested party, even those having no technical background, using only public information; and statistical robustness, supporting drawings where candidates may have distinct probability distributions.
Moreover, it is capable of inviting and engaging as participating stakeholders the main interested parties of a legal process, in a way that promotes process transparency, public trust and institutional resilience.
An open-source implementation is also provided as supplementary material..

   
\end{abstract}

\mbox{} 

 \begin{flushright} 
 \noindent 
 \textit{The function of the legal system is the... congruent  \\ generalization of normative behavior expectations.} \\  
 Niklas Luhmann (1985), A Sociological Theory of Law. 
 \vspace{2mm} 
 \end{flushright}

\section{Introduction} 
\label{sec:intro}

Randomization procedures are routinely used in the design of scientific experiments, in medical trials, and in the operation of legal systems. 
Its use is motivated by the capacity to shield processes against the possibility of all sorts of information biases, extraneous influences, illegitimate interference or spurious manipulations, independently from intention, concealment, or manifestation.
Indeed, in the general framework of randomized experiments \cite[p.340-348]{Pearl:2009}, this shielding is accomplished via a composition of two operations: intervention and randomization.
In medical trials, for example, the intervention is realized when a set of participants, called the experiment group, is treated with the new drug that needs to be tested.
The remaining participants, collectively called the control group, may then receive no intervention, or simply a placebo (aiming to distinguish eventual psychological effects created by the test itself).
However, for a variety of reasons, the decision to which group a patient is assigned may be biased by those conducting the trials; analogously, knowledge about the assignment process itself may allow a participant to infer its corresponding group.
Hence, aiming to produce reliable results, the patient-group allocation should be unpredictable for all entities involved, i.e., it should be realized via randomization.

In the specific context of legal procedures, randomization is employed by many countries as a tool to avoid (the perception of) biased decisions.
Examples include the selection of jurors \cite{random-juror:2002} and judges \cite{random-judges:2012}, in which the main goal is to guarantee that each candidate has a pre-defined (not necessarily uniform) probability of being picked.
In this scenario, though, randomization comes with two additional requirements: auditability by design and active social engagement.
More precisely, auditability by design improves the trust in the system.
Hence, it can avoid suspicions commonly raised when statistical deviations are observed in a non-auditable random procedure \cite{Marcondes:2019}, even if such biases are not the result of ill-intent.
Meanwhile, an active, self-reflective and well-coordinated participation by pertinent members of a community can result in more engagement and inclusiveness, relevant aspects of social practices that also apply to the legal system 
 \cite{Stern:2018,Wenger:1998}.
Combined, such requirements can help legal systems to achieve an important goal: to ensure that its norms (expressed as laws, procedures and regulations) are well understood, recognized, valued  \cite{Luhmann:1985,Luhmann:1989,Stern:2018}.

The scientific understanding of randomization procedures is linked to development of mathematical statistics and cryptography (for a historical overview, see \cite{Marcondes:2019,Stern:2008}).  
After all, randomness is a critical component of any cryptographic solutions involving secret keys, leading to the need of tools for generating (pseudo)random numbers and for statistically assessing their suitability \cite{Hammersley:1964,nist-prng-statistic:2010,nist-prng:2015,Ripley:1987}.
Ensuring that the randomness generator can be audited by anyone, on the other hand, is a more challenging issue.
Some solutions in the literature rely on on the concept of ``open hardware'', so anyone with technical enough background can (at a given time) examine and evaluate the internal circuit and components of the hardware responsible for generating randomness \cite{random-auditable-hardware:2016}. 
There are also proposals that rely on distributed solutions that are expected to generate randomness as part of its regular operation, such as cryptocurrencies \cite{Saa:2019}, thus facilitating auditing by non-technicians.
One drawback of this approach, however, is that the resulting application's security and availability may be affected by external events unrelated to the application itself, but typical of the underlying solution (e.g., forks, implementation bugs, or collusion attacks) \cite{challenges-cryptocurrencies-forks+bugs:2015,challenges-cryptocurrencies-bugs:2017}.
Traditionally, auditability of random results has been discussed by protocols for online games involving chance \cite{gambling-bit-commitment:1997,lotteries-commitment:2004,securetcg:2014}.
Nevertheless, the requirements in those applications are commonly different from the drawing in legal procedures, in particular due to the asymmetry of participants (e.g., the casino owner vs. the players) and the focus on strictly uniform probability distributions.

In this article, we describe an auditable random drawing protocol that combines social engagement and support for multiple probability distributions.
Therefore, it is particularly suited for the context of legal procedures.
The solution builds upon the properties of hash-based bit-commitment mechanisms \cite{bit-commitment-prf-naor:1990}, so it can be executed quite efficiently.
In addition, the scheme's security does not rely on any third-party system; instead, its fairness is assured as long as at least one stakeholder participating in the drawing correctly executes the protocol.
At the same time, auditability in the system requires no software or hardware analysis, but only the set of messages publicly exchanged among the stakeholders.

Section \ref{sec:randomization-legal} discusses the use and importance of randomization in  legal procedure, using the Brazilian legal system as an example. 
Section \ref{sec:proposal} presents the proposed protocols in detail. 
Section \ref{sec:security} analyzes the different security aspects of the protocol. 
Section \ref{sec:application} presents some examples of the protocols developed in this article applied to typical operations in the legal system.  
Section \ref{sec:conclusions} presents our final considerations.

\section{The role of randomization in legal systems: the case of Brazil} 
\label{sec:randomization-legal}

The consolidation of modern democracies presupposes the separation of powers.
In particular, an independent judicial branch is commonly seen as essential to properly check an excessive or abusive exercise of power by the other branches of government  \cite{Hamilton:1788}.
At the same time, such independence promotes the impartiality of judges, i.e., the absence of personal interests or preferences in a trial \cite{Montesquieu:1758}.
The importance of a impartial judiciary is such that it was elevated to the status of a fundamental guarantee by the Universal Declaration of Human Rights, whose Article 10 states that ``\textit{Everyone is entitled in full equality to a fair and public hearing by an independent and impartial tribunal, in the determination of his rights and obligations and of any criminal charge against him}'' \cite{human-rights:1948}.

In Brazil, impartiality is closely related to the guarantee of the natural judge, i.e., everyone shall be entitled to be judged by a court and a judge previously designated in accordance with the law. 
In this context, it is important to ensure a random distribution of the lawsuits among the several judges and/or justices that compose the courts of first instance, the tribunals of second instance and the supreme courts. 
Accordingly, apart from exceptions established by law, the distribution of cases must be randomized, so there is no prior designation of the judge and all members of the court receive a similar number of cases.
In particular, a random distribution is important in repetitive demands for which there are different interpretations of the same law by each judge. 
After all, impartiality would be at risk if a plaintiff could somehow manipulate the distribution criteria aiming to have a case attributed to a judge who ruled it favorably. 
%

Recognizing the importance of randomization in the legal system, the Brazilian Code of Civil Procedure establishes that `` \textit{distribution [of cases] will be made according to the internal rules of procedure of the court, observing the alternation, electronic draw and publicity}" \cite[Art. 930]{codigo-civil-br:2018}.
In the Federal Supreme Court, this is accomplished via a computerized system that is expected to be public and have its data accessible to interested parties \cite[Art. 66]{STF:2020}.
Such publicity is in accordance with the Brazilian Access to Information Act (AIA) \cite{lei-acesso-info-br:2011}, which stipulates as a rule the access to all information and data held by the Government.
However, the computer system responsible for distributing lawsuits has never had its details publicized, and the successive requests for doing so have been denied by the supreme court \cite{Rover:2018}.
One of the main arguments for the refusal is that the specification and source code employed by this system should be covered by secrecy, evoking one caveat contained in the Brazilian AIA \cite[Art. 22]{lei-acesso-info-br:2011}: ``\textit{The provisions of this Law do not exclude the other legal hypotheses of secrecy and secrecy of justice or the hypotheses of industrial secrecy arising from the direct exploitation of economic activity by the State or by a natural person or private entity that has any link with the public authorities}''.
In practice, however, such secrecy creates a ``security through obscurity" system, which has been considered a poor practice by security practitioners for more than 100 years due to its inherent lack of auditability \cite{Kerckhoffs-security-by-obscurity:1883}.
Hence, there are no technical grounds to support secrecy of the algorithms and source code employed, while the legal grounds are still a matter of dispute.

Unfortunately, until this controversy is resolved (e.g., by the bill of law 8503/2017, which compels the removal of such secrecy \cite{Rodrigues:2017}), the system will remain unable to provide enough transparency to assuage eventual suspicion and distrust, even if unjustified.
This issue is especially troublesome when we consider that the Supreme Court is often called to decide delicate questions that are subject of heated debate in the society at large.
In such cases, any distrust motivated by security by obscurity may spill over other social systems, spreading institutional discredit to a much wider scope and, in so doing, potentially threaten social harmony or stability \cite{Luhmann:1989}.

Such concerns motivate the development of proposals following a \textit{security by design} concept, which implies that the system's security does not depend on the secrecy of its implementation or of its components \cite[Sec. 2.4]{NIST-server-security:2008}.
In the specific case of Brazil, this approach is expected to avoid any clashes with the principles of publicity imposed by the Federal Constitution, the Code of Civil Procedure and the AIA.
The main goal of the remainder of this article is to show that it is possible to specify and implement such a solution having transparency and auditability at its core.

\section{Auditable random draw}
\label{sec:proposal}

In this section, we describe the process of randomly drawing some entity among a list of eligible candidates.
The proposed protocols build upon the ideas originally discussed by M. Blum for solving the ``Coin-flipping by telephone'' problem \cite{Blum:1983,Brassard:1988}, where two mutually untrusted parties play a virtual coin tossing game: after each player chooses ``heads'' or ``tails'', an outcome is randomly drawn in such a manner that both players can verify the fairness of the result (i.e., in this case, that each one had a 50\% chance of winning).
Basically, the solution employs a commit-and-reveal scheme \cite{bit-commitment-prf-naor:1990}, leading to a protocol that is general enough to be applied in a variety of applications. 
Indeed, it has been traditionally employed in protocols for online gambling  \cite{gambling-bit-commitment:1997} and peer-to-peer card games \cite{securetcg:2014}.
In this article, though, we focus specifically on the context of legal cases, assuming that entities like judge, juror(s), rapporteur, or the court itself must be selected at random in a judicial proceeding.

We discuss two main protocols: one version where a single drawing is required for a given proceeding, and an extension that optimizes latency and bandwidth usage in scenarios where multiple entities must be simultaneously drawn for the same or for several proceedings.
We also discuss some possible protocol variants, as well as how the described schemes could be instantiated in for handling real-world judicial proceedings.

%

%


\subsection{Preliminaries: formal description and notation}
\label{sec:proposal-notation}

For convenience to the reader, Table~\ref{tab:symbols} lists the general notation adopted hereinafter.

\begin{table}[ht]
\centering
\caption{General notation}
\label{tab:symbols}
\small
\begin{tabular}{c|l}\hline
Symbol                  &   Definition  \\\hline
$\seclevel$             &   System's security level  \\
$x \samples X$          &   Uniform sampling of an element $x$ from space $X$   \\
$\len{Y}$               &   Number of elements in a set or list $Y$   \\
$\draw$                 &   A random drawing procedure \\
$\drawSet = \{\draw_0,\ldots\}$              
                        &   A list of drawing procedures  \\
$\procStakeholders = \{\stakeholder_0,\ldots\}$     
                        &   Set of stakeholders $\stakeholder_j$ participating in drawing procedure $\draw$ \\
$\procEligible =  \{\eligible_0,\ldots\}$
                        &   Ordered list of eligible candidates $\eligible_j$ in drawing procedure $\draw$ \\
$\drawid$               &   Unique identifier of a drawing procedure $\draw$    \\
$\procinfo$             &   Any metadata related to drawing procedure $\draw$  \\
$\share$                &   A stakeholder's contribution to the random draw \\
$\commit$               &   Commitment to the contribution $\share$ in a given drawing  \\
$\mask$                 &   A random masking value: hides contribution $\share$ in commitment $\commit$     \\
$\drawnValue$           &   The result of the random draw   \\
$\pk, \sk$              &   An entity's public and private keys, respectively  \\
$\hash(M)$              &   Hash of an arbitrary message $M$   \\
$\sig$                  &   A digital signature   \\
$\sign{\sk}{M}$         &   Signing message $M$ using private key $\sk$ \\
$\verif{\pk}{M}{\sig}$  &   Verification of signature $\sig$ on message $M$, using public key $\pk$   \\

\hline
\end{tabular}
\end{table}

In the described protocols, we consider that each drawing procedure $\draw$ can be represented by the set of fields $\{\drawid, \procStakeholders, \procEligible, \procinfo\}$, described as follows:
\begin{itemize}
    \item $\drawid$ (mandatory): a unique identifier for the drawing procedure. 
    In particular, when a drawing is associated with a proceeding whose unique identifier is $\procid$, one might simply make $\drawid = \procid \cat \cnt$, where $\cat$ denotes concatenation (using a suitable, reserved character) and $\cnt$ is a counter for the number of the drawing inside that proceeding.
    For example, suppose that a proceeding's identifier is $\procid = $ {\tt 123.456-7}, and that a random draw is required for defining its judge.
    This first drawing could then be identified as $\drawid = $ {\tt 123.456-7\#0}.
    
    \item $\procStakeholders$ (mandatory): the set of all stakeholders $\stakeholder_j$ (where $0 \leqslant j < \len{\procStakeholders}$) that must participate in the random draw as witnesses of its fairness.
    This set may contain any number of interested parties, which may be either proceeding-specific (e.g., defense lawyer, prosecutor, and judge) or more general (e.g., Ministry of Justice, Supreme Court, and bar council).
    Each interested party must be identified by a public key, so their corresponding digital signatures can be verified during the protocol's execution. 
    Without loss of generality, we assume that the public key $\pk_{\stakeholder_j}$ of each interested party $\stakeholder_j \in \procStakeholders$ is part of a digital certificate issued by trusted Certificate Authority (CA), so that certificate's fingerprint can be used as an unambiguous identifier.

    \item $\procEligible$ (mandatory): the list of all candidates $\eligible_j$ (where $0 \leqslant j < \len{\procEligible}$) that are eligible to be randomly drawn.
    For example, it might refer to all judges that are eligible for the proceeding, excluding entities with conflict of interest; it may also including duplicates, aiming to handle non-uniform probability distributions (see Section \ref{sec:proposal-non-uniform} for details).
    The identification of each candidate and their order in the list must be unequivocal. 
    This can be accomplished, for example, by means of a list containing their corresponding social security numbers, functional identifiers, or digital certificate fingerprints, sorted in lexicographic order.
    
    \item $\procinfo$ (optional): represents all relevant metadata about the drawing in a human-readable form. 
    This field might include, for example, the proceeding title, class, subject, and last modification date. 
    This field is left as optional in the protocol because, if a reliable source is available, such metadata can be obtained from $\drawid$ itself.
    
\end{itemize}

We denote by $\hash(M)$ the application of a hash-function $\hash : \{0,1\}^* \rightarrow \{0,1\}^h$ over the arbitrary-length input $M$.
In the protocols hereby described, hash functions are employed in the construction of a commitment mechanism \cite{bit-commitment-prf-naor:1990}: after computing and revealing $\hash(M)$, an entity becomes ``committed'' to $M$, since it is computationally hard to find $M' \neq M$ such that $\hash(M') = \hash(M)$ for a secure hash function; at the same time, one-way property of the hash-function prevents anyone from learning the value of $M$ until it is deliberately disclosed.
We also assume that $\hash$ follows a fairly uniform distribution in $\{0,1\}^h$, which is standard for secure hash functions.
Standardized algorithms believed to provide such properties include instances from the SHA-2 \cite{sha2:2015} family.

We write $\sign{\sk}{M}$ to denote the computation of a digital signature of input $M$ using the private key $\sk$, giving as output a signature $\sig$.
The corresponding signature verification procedure under public key $\pk$ is then denoted $\verif{\pk}{M}{\sig}$.
We assume that a standardized algorithm is employed for this purpose, such as ECDSA or EdDSA \cite{nist-dss:2019}.

For all algorithms employed, we assume a security level $\seclevel \geq 128$ bits, as it is usual in modern systems \cite{security-levels-nist:2019}.

\subsection{Single random draw}
\label{sec:proposal-one-proc}

\begin{figure}[tb!]
    \centering
    \includegraphics[trim={3pt 4pt 3pt 6pt},clip,width=0.62\textwidth]{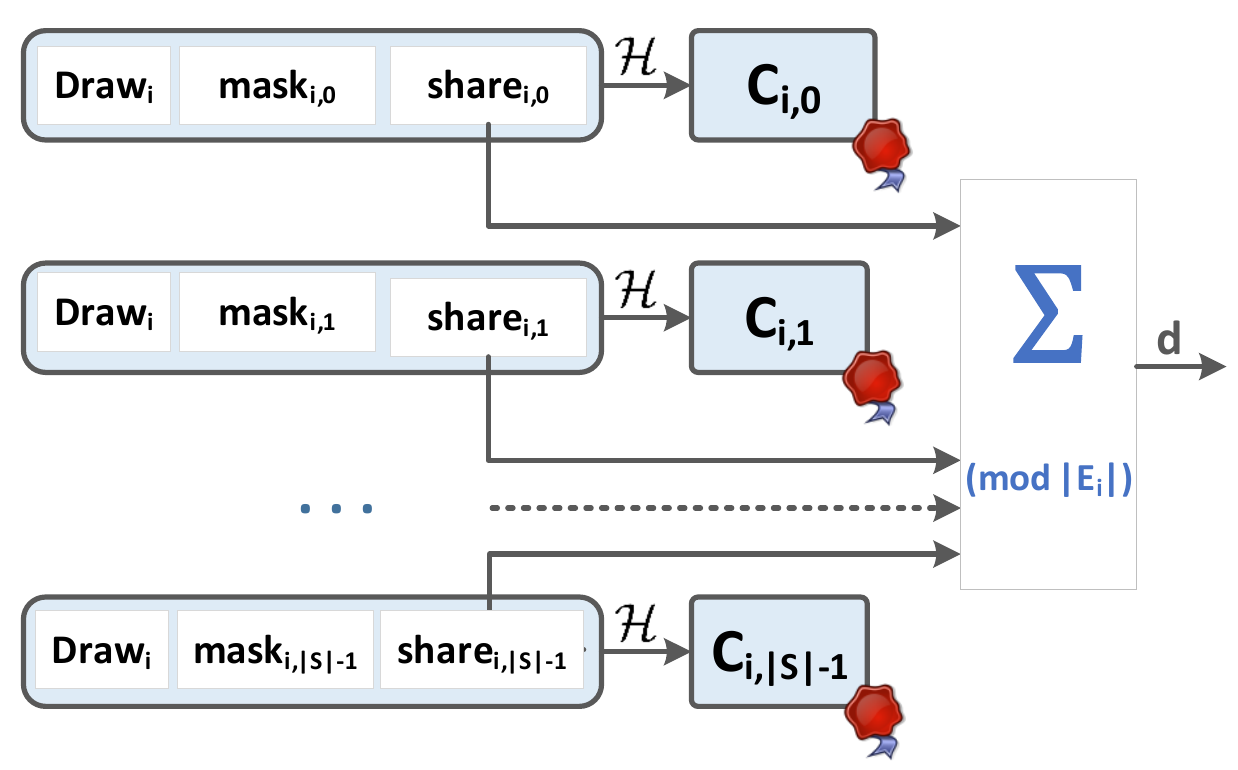}
    \caption{Auditable random draw procedure.}
    \label{fig:one-draw}
\end{figure}

Let $\draw_i = \{\drawid_i, \procStakeholders_i, \procEligible_i, \procinfo_i\}$ represent a random drawing procedure performed by stakeholders $\procStakeholders_i$.
To pick a random candidate from $\procEligible_i$, each stakeholder $\stakeholder_j \in \procStakeholders_i$ engages in a two-phase procedure, described in what follows and illustrated in Figure \ref{fig:one-draw}.

\subsubsection{Commitment phase}
\label{sec:proposal-one-proc-commit}

Firstly, $\stakeholder_j$ generates a random masking value $\mask_{i,\:j} \samples \bitset{\seclevel}$ for security level $\seclevel$.
In addition, $\stakeholder_j$ picks a random value $\share_{i,\:j}$ satisfying $0 \leqslant \share_{i,\:j} \leqslant \len{\procEligible_i}$, which will later be used as that stakeholder's contribution to the random draw.
%
%
We note that, as long as both $\mask_{i,\:j}$ and $\share_{i,\:j}$ are kept secret and can be considered unpredictable, their values could be picked arbitrarily by $\stakeholder_j$ or computed using a suitable random number generator \cite{nist-prng-statistic:2010,nist-prng:2015}.

Subsequently, each stakeholder $\stakeholder_j$ computes its own commitment $\commit_{i,\:j} \gets \hash(\draw_i,  \mask_{i,\:j}, \share_{i,\:j})$ by applying the hash function $\hash$ on the drawing data $\draw_i$ (common to all parties), on the masking value $\mask_{i,\:j}$, and on its random contribution $\share_{i,\:j}$.
With this approach, the potentially low-entropy hash input $\share_{i,\:j}$ cannot be guessed from $\commit_{i,\:j}$, since it is combined with the high-entropy masking value $\mask_{i,\:j}$ \cite{bit-commitment-prf-naor:1990}.

Finally, $\stakeholder_j$ signs a message containing the commitment $\commit_{i,\:j}$ and the drawing data $\draw_i$, using the private key $\sk_j$.
The digital signature generated in this manner, $\sig_{i,\:j} \gets \sign{\sk_j}{\{\draw_i, \commit_{i,\:j}\}}$, provides authenticity and non-repudiation to the commitment sent by $\stakeholder_j$, which allows latter auditing.
Finally, $\stakeholder_{j}$ broadcasts a message containing $\{\draw_i, \commit_{i,\:j}, \sig_{i,\:j}\}$ to all other stakeholders $\stakeholder_{j' \neq j}$.

\subsubsection{Reveal phase}
\label{sec:proposal-one-proc-reveal}

Upon reception of a commitment $\commit_{i,\:j'}$, each stakeholder $\stakeholder_j$ checks the corresponding signature by running the verification algorithm $\verif{\pk_{j'}}{\{\draw_i, \commit_{i,\:j'}\}}{\sig_{i,\:j'}}$. 
Only after all commitments $\commit_{i,\:j' \neq j}$ are received and their signatures are correctly verified, stakeholder $\stakeholder_j$ reveals the pair $\{\mask_{i,\:j}, \share_{i,\:j}\}$ to all of its peers.
Note that it is not necessary to digitally sign the message revealed in this manner, since $\{\mask_{i,\:j}, \share_{i,\:j}\}$ was indirectly signed when computing $\sig_{i,\:j}$: to verify its validity, it is enough to check that $\commit_{i,\:j} \iseq \hash(\draw_i,  \mask_{i,\:j} , \share_{i,\:j})$ holds true.

Using the random contributions $\share_{i,\:j}$ from all stakeholders, the result of the random draw is $\drawnValue = (\sum_{j=0}^{\len{\procStakeholders_i}-1} \share_{i,\:j}) \bmod \len{\procEligible_i}$.
The drawn candidate is then set to $\eligible_{\drawnValue}$, following the original order of candidates from $\procEligible_i$.
This approach ensures that every candidate $\eligible_j$ has the same probability of being drawn because, if at least one stakeholder $\stakeholder_j$ picks $\share_{i,\:j}$ uniformly at random in $[0 , \len{\procEligible_i}[$, the resulting sum will also be uniformly distributed in the same interval \cite{uniform-distribution-sum-modulo:1993}, independently of collusion among other parties.
In addition, any entity is capable of auditing the drawing by: (1) verifying the digital signatures on the revealed values; (2) recomputing $\drawnValue$ independently; and (3) comparing the obtained $\drawnValue$ with the value reported by the stakeholders that participated in the drawing.

\subsection{Multiple random draws by the same stakeholders}
\label{sec:proposal-n-proc}


The process described in Section~\ref{sec:proposal-one-proc} can be extended to enable multiple random draws to be executed by a group of stakeholders $\procStakeholders$ with a single commit-and-reveal procedure.
This extension is discussed in what follows.

\subsubsection{Commitment phase}
\label{sec:proposal-n-proc-commit}

\begin{figure}[tb!]
    \centering
    \includegraphics[trim={19pt 21pt 21pt 20pt},clip,width=\textwidth]{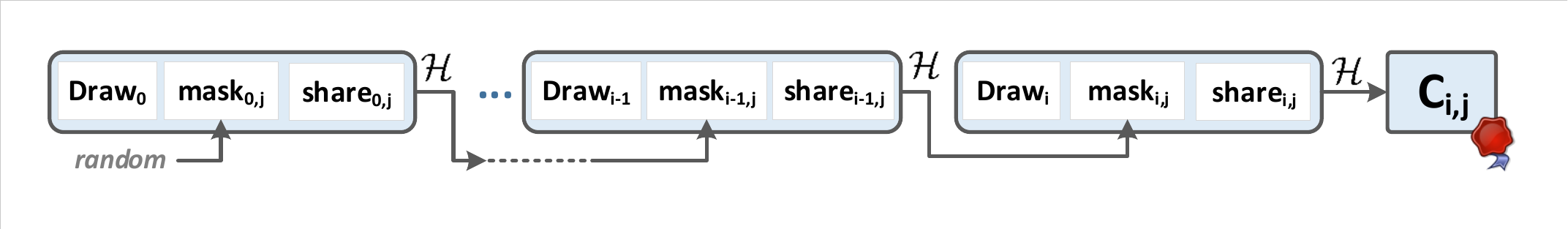}
    \caption{Chaining structure enabling multiple random draws from a single commitment.}
    \label{fig:multi-draw-blockchain}
\end{figure}

Let $\drawSet = \{\draw_i\}$ (for $i \geqslant 0$) be a list of random draws $\{\drawid_i, \procStakeholders, \procEligible_i, \procinfo_i\}$ that share the same set of stakeholders $\procStakeholders$ and that are ordered according to some rule (e.g., following the lexicographic order of $\drawid_i$).
%
%
Similarly to the single-drawing case, each stakeholder $\stakeholder_j \in \procStakeholders_i$ starts by picking a random $\mask_{0,\:j} \samples \bitset{\seclevel}$.
In addition, $\stakeholder_j$ picks one random $\share_{i,\:j}$ for each $\draw_i \in \drawSet$, each of which satisfying $0 \leqslant \share_{i,\:j} \leqslant \len{\procEligible_i}$ for the corresponding $\procEligible_i$.
The $\drawSet$ commitments from $\stakeholder_j$ are then obtained iteratively: first, by making $\commit_{0,\:j} \gets \hash(\draw_0,  \mask_{0,\:j}, \share_{0,\:j})$; the subsequent $\commit_{i,\:j}$ for $i \geqslant 1$ are then computed as $\commit_{i,\:j} \gets \hash(\draw_i, \mask_{i,\:j}, \share_{i,\:j})$, where $\mask_{i,\:j} = \commit_{i-1,\:j}$.
The resulting data structure is illustrated in Figure \ref{fig:multi-draw-blockchain}.
Finally, the last commitment $\commit_{\len{\drawSet},\:j}$ computed in this manner is signed and broadcast to all stakeholders.

\subsubsection{Reveal phase}
\label{sec:proposal-n-proc-reveal}

After $\stakeholder_j$ receives and validates all commitments $\commit_{\len{\drawSet},\:j' \neq j}$ from its peers, it broadcasts $\mask_{0,\:j}$ together with all picked values of $\share_{i,\:j}$ (for $i \geqslant 0$).
This allows any entity, including stakeholders, to verify that the signed commitment $\commit_{\len{\drawSet},\:j}$ originally provided by $\stakeholder_j$ was indeed built from $\mask_{0,\:j}$ and the set of disclosed $\share_{i,\:j}$: it suffices to reproduce the aforementioned procedure that, supposedly, was followed by $\stakeholder_j$ when computing each $\commit_{i,\:j}$.
If such verification holds true for all commitments, each random draw $\drawnValue_i$ is once again computed as $\drawnValue_i = (\sum_{j=0}^{\len{\procStakeholders_i}-1} \share_{i,\:j}) \bmod \len{\procEligible_i}$ for each $i \geqslant 0$.
Once again, the fairness of the drawing procedure can be audited by independent entities, who are able to verify that $\drawnValue$ was computed from the signed commitments.

\subsection{Handling non-uniform drawing probabilities}
\label{sec:proposal-non-uniform}

Many real-world random drawing applications require that $n$ eligible candidates in a list $\procEligible$ have the same probability of being drawn, that is, a uniform probability distribution.
In this case, the ordered list $\procEligible =  \{\eligible_0,\ldots\,\eligible_{n-1}\}$ would contain only distinct identifiers, one per candidate $\eligible_j$.

Nevertheless, there are situations in which the $n$ eligible candidates must be selected according to a non-uniform probability distribution $\prob{0},\prob{1} \ldots \prob{n-1}$, where $\prob{j} \geqslant 0$ and $\sum \prob{j} =1$.
For example, in the context of legal proceedings, some publicly available and law-abiding rules may dictate that the judge for a given case should be picked with higher or lower probability depending on well-established methodologies and criteria.
For example, these criteria may include judges' current workloads, case complexities or legal specialty areas, among other.
These probability distributions may even be adjusted along the time aiming to make the judges' loads converge, in the long run, to a targeted equilibrium goal.
Some statistical methods for calculating, calibrating and adjusting such non-uniform distributions are discussed in  \cite{Fossaluza:2015,Lauretto:2012,Lauretto:2017}.
  
A standard technique for handling non-uniform  probability distributions consists in repeating the identifier of every candidate $\eligible_j$ proportionally to $\prob{j}$.  
The case in which probabilities are expressed as fractions with a common denominator, $\prob{j}=a_j/b$ is simple to handle: 
we only have to build 
$\procEligible$ as a $b$-long list where the identifier for each candidate  $\eligible_j$ appears (e.g., contiguously) a total of $a_j$ times.
For example, if we need a random draw among 4 candidates with probability distributions $\{1/10, 2/10, 3/10, 4/10\}$, where $b=10$, we would have 
$\procEligible = \{\eligible_0,  \eligible_1, \eligible_1,  \eligible_2,  \eligible_2, \eligible_2,, \eligible_3, \eligible_3,  \eligible_3, \eligible_3 \}$.  
Taking as common denominator a larger integer power of ten, i.e. $b=10^k$, allows for a good approximation of any distribution expressed in  decimal form, like a centesimal or a millesimal scale for a common denominator of $b=100$ or $b=1000$.

The case in which probabilities are expressed as fractions in canonical form, $\prob{j}=a_j/b_j$, with no common denominator, is handled as follows: 
(1) compute $\ell\gets \lcm(b_0, b_1,\ldots)$, i.e., the lowest common multiple of the fractions' denominators, $b_j$; and 
(2) build $\procEligible$ as a $\ell$-long list where the identifier for each candidate  $\eligible_j$ appears (e.g., contiguously) a total of $\ell\!\cdot\!a_j/b_j$ times.  
For example, if we need a random draw among 4 candidates with probability distributions 
$\{1/6, 1/4, 1/4, 1/3 \}$, 
then we would have 
$\ell\gets \lcm(3,4,6)=12$, and 
$\procEligible = 
\{\eligible_0, \eligible_0,  \eligible_1, \eligible_1, \eligible_1, 
\eligible_2, \eligible_2, \eligible_2, 
\eligible_3, \eligible_3, \eligible_3, \eligible_3 \}$.

Despite repetitions in the list $\procEligible$, we note that the computational representation of $\procEligible$ can remain quite compact: by representing each candidate by the pair $(\eligible_j,\prob{j})$, no actual identifier repetition is necessary.

\subsection{A possible variant, aimed at better bandwidth efficiency}
\label{sec:proposal-variant}

A slightly modified version of the described protocols can be employed aiming to save some bandwidth during the reveal phase.
This variation consists in use the masking values $\mask_{i,\:j}$ directly as source of randomness instead of relying on the additional random values of $\share_{i,\:j}$.
For the single drawing procedure from Section \ref{sec:proposal-one-proc}, this means that $\drawnValue$ would be computed by adding up $\mask_{i,\:j}$, i.e., as $\drawnValue = (\sum_{j=0}^{\len{\procStakeholders_i}-1} \mask_{i,\:j}) \bmod \len{\procEligible_i}$.
In this case, $\share_{i,\:j}$ itself could be omitted from the protocol, and only $\mask_{i,\:j}$ would be revealed by the stakeholders to their peers.
In addition, multiple random draws could then be implemented without the chaining structure described in Section \ref{sec:proposal-n-proc}: instead, one could employ a pseudo-random number generator \cite{nist-prng:2015} taking as seed the value of $\drawnValue$ obtained in the single-drawing procedure.

The drawback of this approach is that the distribution of $\drawnValue$ computed in this manner may lead to distortions in the protocol's probability distribution. 
Specifically, the lowest $(2^\seclevel\len{\procStakeholders_i} \bmod \len{\procEligible_i})$ values of $\drawnValue$ would have a favorable probability bias: instead of being selected with probability $1/\len{\procEligible_i}$, their actual chance would be $1/\len{\procEligible_i} + 1/2^\seclevel$.

Notice that such probability issue only arises in this modified protocol when $2^\seclevel\len{\procStakeholders_i} \bmod \len{\procEligible_i} \neq 0$.
In addition, the resulting bias should be negligible whenever $\len{\procEligible_i} \ll 2^\seclevel$, which is likely to be the case in many real-world applications.
For example, one would expect a small $\len{\procEligible_i}$ when the judge for a procedure needs to be randomly drawn according to an uniform distribution.
Nevertheless, $\len{\procEligible_i}$ may grow for supporting arbitrary drawing probabilities associated with each candidate.
Therefore, aiming to ensure the wide applicability of the hereby described protocols, we recommend using $\share_{i,\:j}$ as an additional value in actual implementations.

\section{Security Analysis}
\label{sec:security}

In this section, we analyze the attack surface of the proposed secure drawing mechanism, considering the security properties of its underlying cryptographic primitives.

\subsection{Confidentiality of stakeholders' contributions in the Commitment phase}
\label{sec:security-confident}

Suppose a malicious stakeholder $\stakeholder_a$ is able to learn all contributions $\share_{i,\:j{\neq}a}$ from its peers before sending its own commitment $\commit_{i,\:a} \gets \hash(\draw_i, \mask_{i,\:a}, \share_{i,\:a})$.
In that case, $\stakeholder_a$ can choose the value of $\drawnValue_i$ by picking $\share_{i,\:a}$ accordingly.
The confidentiality of all $\share_{i,\:j}$ in the commitment phase is, thus, critical for the drawing procedure's fairness.

In the described protocol, the confidentiality of every pre-image resistance during the commitment phase is protected by the underlying hash function's pre-image resistance.
Specifically, to obtain $\share_{i,\:j}$, $\stakeholder_a$ would have to find the hash function's input $(\draw_i, \mask_{i,\:j}, \share_{i,\:j})$ from its output $\commit_{i,\:j}$.
This requires guessing $\mask_{i,\:j}$ in the one-draw protocol described in Section \ref{sec:proposal-one-proc}, or $\mask_{0,\:j}$ in the multi-draw protocol from Section \ref{sec:proposal-n-proc}.
As long as such masking values are at least $\seclevel$-bits long and randomly picked, such guessing attempts should be computationally infeasible.

Notice that the confidentiality of every $\share_{i,\:j}$ is relinquished in the reveal phase, when those values are disclosed together with the corresponding $\mask_{i,\:j}$.
At that time, however, it would be computationally hard for $\stakeholder_a$ to modify the already committed $\share_{1,\:a}$, picked before any $\share_{i,\:j{\neq}a}$ was known (see Section \ref{sec:security-modifyCommited}).
Hence, the drawing procedure cannot be manipulated as long as every $\stakeholder_j$ reveal its own $\{\mask_{i,\:j}, \share_{i,\:j}\}$ only after all commitments $\commit_{i,\:j'{\neq}j}$ are received from their peers.

\subsection{Immutability of stakeholders' contributions in the Reveal phase}
\label{sec:security-modifyCommited}

Suppose a malicious stakeholder $\stakeholder_a$ can modify its own $\share_{i,\:a}$ after learning all contributions $\share_{i,\:j{\neq}a}$ from its peers.
In this scenario, similarly to the attack described in Section \ref{sec:security-confident}, $\stakeholder_a$ can pick a modified value $\share_{i,\:a}'$ that leads to the desired value of $\drawnValue_i$.

In the described protocols, such attack is unfeasible as long as a collision-resistant hash function $\hash$ is employed when computing the commitment $\commit_{i,\:a}$.
More precisely, after $\stakeholder_a$ broadcasts its commitment $\commit_{i,\:a} = \hash(\draw_i,  \mask_{i,\:a} , \share_{i,\:a})$, the value of $\{\mask_{i,\:a}', \share_{i,\:a}'\}$ subsequently revealed would only be accepted as valid by its peers if the following collision occurs: $\hash(\draw_i, \mask_{i,\:a}, \share_{i,\:a}) = \hash(\draw_i, \mask_{i,\:a}', \share_{i,\:a}')$.

Notice also that attempts to replace $\commit_{i,\:a}$ itself during the reveal phase would also fail.
After all, stakeholders would not enter the reveal phase until $\commit_{i,\:a}$ is received and its signature is verified.

\subsection{Split decision via duplicated commitments}
\label{sec:security-splitDecision}

A malicious stakeholder $\stakeholder_a$ might decide to send different commitments to different sets of stakeholders, leading to a distinct value of $\drawnValue$ computed in each of them.
The result would be a denial-of-service attack, because there would be no consensus among all stakeholders.
Even though there is no mechanism to prevent such attack, the culprit can be easily identified after the stakeholders compare the received commitments.
The attacker could then be penalized accordingly, and the digitally signed commitments could be used as proof of misbehavior.




\subsection{Collusion resistance}
\label{sec:security-collusion}

As mentioned in Section \ref{sec:proposal-one-proc}, the value of $\drawnValue = (\sum_{j=0}^{\len{\procStakeholders_i}-1} \share_{i,\:j}) \bmod \len{\procEligible_i}$ obtained in the hereby described protocol follows an uniform distribution in $[0 , \len{\procEligible_i}[$ as long as at least one stakeholder $\stakeholder_j$ picks $\share_{i,\:j}$ uniformly at random in $[0 , \len{\procEligible_i}[$ \cite{uniform-distribution-sum-modulo:1993}.
%
Hence, the fairness of the random draw is ensured even if $\len{\procStakeholders_i}-1$ stakeholders collude, e.g., by revealing and/or agreeing on their own contributions $\share_{i,\:j'{\neq}j}$.

We note that, if there is a collusion among all stakeholders (i.e., a consensus), then it is possible to manipulate the drawing procedure while giving auditors a false impression of fairness.
Hence, the choice of a suitable set of stakeholders $\procStakeholders$ is a critical requirement in the system.
In the specific case of drawing a proceeding's judge, meeting such requirement should be quite easy, in particular if opposing parties like the defense lawyer and prosecutor are included as  $\procStakeholders$.

\subsection{Impersonation attacks: commitment replay or forgery}
\label{sec:security-replay}

The successful impersonation of a honest stakeholder $\stakeholder_j$ might lead to a few undesirable situations.
For example, suppose that both the legitimate and a forged/replayed commitment from $\stakeholder_j$ are accepted as valid in a random draw, 
Since the resulting duplication would be indistinguishable from the denial of service attack described in Section \ref{sec:security-splitDecision}, $\stakeholder_j$ might be unjustly accused of misbehavior.
As another example, suppose that $n$ stakeholders in collusion gather forged/replayed commitments from all of the remaining $\len{\procStakeholders_i}-n$ stakeholders that would participate in a drawing.
In that case, auditors could be tricked into believing that a given drawing result was fair, when it was actually manipulated by the colluding parties.

To prevent such attacks, two mechanisms are employed in the hereby described protocols.
First, to prevent forgery, all stakeholders must be unequivocally identified (e.g., by their digital certificates) and their commitments must be signed using a secure digital signature algorithm.
Second, to prevent replay attacks, every random draw procedure $\draw_i$ includes a unique identifier; hence, a commitment $\commit_{i,\:j}$ for $\draw_i$ would not be mistakenly accepted as valid in another drawing procedure $\draw_{i'{\neq}i}$.


\subsection{Denial to reveal}
\label{sec:security-deny}

Any malicious stakeholder $\stakeholder_a$ can engage in a denial of service attack by refusing to provide either $\{\commit_{i,\:a}, \sig_{i,\:a}\}$ or $\{\mask_{i,\:a}, \share_{i,\:a}\}$, preventing the completion of the protocol's execution.
Even though there are no mechanisms to prevent such attacks from occurring, the non-compliant parties can be easily identified in the protocol.
Hence, adequate measures can be taken in response, depending on the target scenario.
For example, if the contribution from $\stakeholder_a$ is not mandatory, then the drawing procedure could be restarted after $\stakeholder_a$ is removed from $\procStakeholders_i$.

\subsection{Dealing with an untrustworthy server}
\label{sec:security-server}

The described protocol requires $\stakeholder_j$ to broadcast $\{\commit_{i,\:j}, \sig_{i,\:j}\}$ (in the commit phase) and $\{\mask_{i,\:j}, \share_{i,\:j}\}$ (in the reveal phase).
Such broadcasts can be performed either directly, using the stakeholders' network addresses, or with the aid of an intermediate server.
One benefit of the latter approach, though, is that each $\stakeholder_j$ would need to send a single message to the server, rather than learning its peers' addresses and sending one individual message to each peer.
Hence, for better efficiency, such a server-based architecture may be preferred in actual deployments
Meanwhile, security-wise, there would be no impact in terms of security: even if the server is untrustworthy, it would be unable to forge or modify any of the exchanged messages because they are all messages signed by the corresponding stakeholders.

The main caveat in a server-based architecture is that third parties interested in auditing the drawing result should not blindly trust the data provided by the server.
The reason is that the server could collude with a a malicious stakeholder $\stakeholder_a$ for replacing the latter's (signed) contribution in the drawing and, thus, manipulate its result from that auditor's perspective.
Hence, auditors should always confirm that any data provided by the intermediate server matches the messages actually seen by all stakeholders.
Notice that such confirmation allows auditors not only to avoid tampering attempts, but also enables the identification of the malicious stakeholder(s) behind this attempt: after all, the auditor would observe two distinct commitments $\commit_{i,\:a}$ and $\commit^{'}_{i,\:a}$ signed by $\stakeholder_a$ for the same drawing $\draw_i$, a situation that should never occur in a regular protocol execution.
Actually, this very possibility of identifying tampering attempts should dissuade stakeholders from colluding with the intermediate server.

\section{Implementation}
\label{sec:application}

We have developed a simple Java library that implements all the steps of the protocols described in Sections \ref{sec:proposal-one-proc} and \ref{sec:proposal-n-proc}.
The source code is available under the MIT License at \urls{https://doi.org/10.24433/CO.6108166.v1}, so it can be freely adapted for fitting the needs of real-world implementation.
It also includes routines for performing the functional testing of the protocol's main routines (a reproducible run is made available).

\begin{figure}[tb!]
    \centering
    \includegraphics[trim={30pt 24pt 32pt 23pt},clip,width=1.01\textwidth]{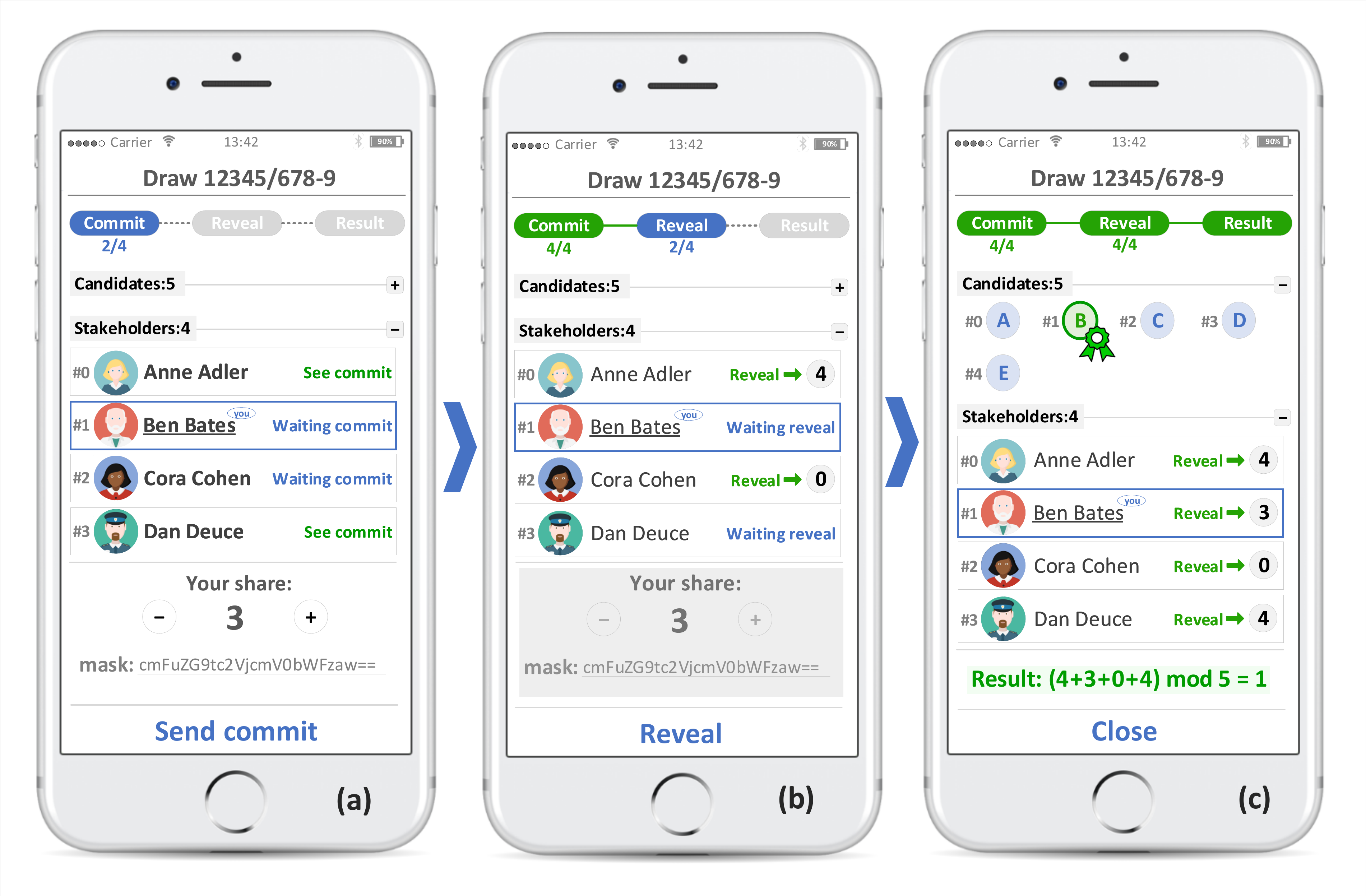}
    \caption{Graphical interface for the described protocol's proof-of-concept implementation: (a) commit phase; (b) reveal phase; (c) end of the protocol, with one out of five candidates being randomly drawn by four stakeholders.}
    \label{fig:poc+screen}
\end{figure}

The provided code does not include a graphical interface, since its details would depend on the actual platform (e.g., desktops, mobile phones or dedicated hardware) and also on the details of the scenario (e.g., usual number of stakeholders, and whether or not non-uniform probability distributions are required).
We are currently implementing a prototype mobile application that uses an intermediate server for facilitating the communication among peers.
Figure \ref{fig:poc+screen} illustrates the graphical interface expected for this proof-of-concept.
Specifically, it shows the look-and-feel for mobile users during: 

\begin{enumerate}[label=(\alph*)]
    \item The commit phase, when 4 stakeholders must send their signed commitments. At the moment shown in the interface, only two of them (namely, \#0 and \#3) were received by the stakeholder (whose identifier is \#1) ; 
    
    \item The reveal phase, starting after all commitments are received, when each stakeholder reveals its own values of $\share$ and $\mask$. The interface shows that two stakeholders (namely, \#0 and \#3) have revealed valid data.
    
    \item The completion of the protocol, when one of the eligible candidates (namely, \#1) is picked with uniform probability based on the stakeholders' contributions.

\end{enumerate}

\section{Final Considerations}
\label{sec:conclusions}

In this article, we describe a collaborative random drawing protocol with arbitrary probability distributions and whose fairness can be audited by any interested party (including non-technicians).
The scheme follows a \emph{security-by-design} best practice, contrasting with technically unsound approaches based on \emph{security-by-obscurity}.
In addition, it is designed to allow and invite the active participation of any number of stakeholders or their representatives.
This active engagement of interested parties and social organizations is intended to foster trust and confidence in the legal processes.
Indirectly, it should also strengthen the institutions that compose a truly autonomous Legal System, enhancing their harmonious relations with other branches of government and, in this way, promoting social peace.

\subsubsection*{Acknowledgements and Funding} 
 
This work was supported by:
Ripple's University Blockchain Research Initiative; 
CNPq (Brazilian National Council for Scientific and Technological Development -- grants PQ 307648/2018-4 and 301198/2017-9); and 
FAPESP (S\~{a}o Paulo Research Foundation, grants CEPID-CeMEAI 2013/07375-0 and CEPID-Shell-RCGI 2014/50279-4). 
The authors are grateful for early conversations with Julio Adolfo Zucon Trecenti from ABJ (Brazilian Jurimetrics Association), and for the mobile interface design conceived by Giovanni A. dos Santos and Joao Paulo A. S. E. Lins.


\bibliographystyle{plain}
\bibliography{refs}

\end{document}